\documentclass[12pt]{article}
\usepackage{amsmath}
\usepackage{amsfonts}
\usepackage{tikz}
\usepackage{geometry}
\usepackage{bm}
\usepackage{setspace}
\usepackage{mathtools}
\renewcommand\Im{\operatorname{Im}}
\renewcommand\Re{\operatorname{Re}}

\newcommand{\Li}{\operatorname{Li}}
\newcommand{\bmid}{\;\ifnum\currentgrouptype=16 \middle\fi|\;}

\DeclarePairedDelimiter\abs{\lvert}{\rvert}
\DeclarePairedDelimiter\norm{\lVert}{\rVert}

\makeatletter
\let\oldabs\abs
\def\abs{\@ifstar{\oldabs}{\oldabs*}}

\let\oldnorm\norm
\def\norm{\@ifstar{\oldnorm}{\oldnorm*}}
\makeatother

\numberwithin{equation}{section}

\usepackage[toc,page]{appendix}

\begin{document}

\begin{titlepage}
\samepage{
\setcounter{page}{0}
 \rightline{ }
\vfill
\begin{center}
\vspace{1.2in}
    {\Large \bf Shift-Symmetries at Higher Order\\} 
\vfill
\vspace{0.2in}
   {\large Steven Abel\footnote{E-mail address:
      {\tt s.a.abel@durham.ac.uk}} \, {\it and} ~
       Richard J. Stewart\footnote{E-mail address:
      {\tt richard.stewart@durham.ac.uk}}\\}
\vspace{0.30in}
   {\it 
     IPPP and  Department of Mathematical Sciences, \\ Durham University, Durham, DH1 3LE, UK\\ }  
\end{center}
\vfill
\begin{abstract}
  {\rm  
\noindent The fate of shift-symmetries in effective string models is considered beyond tree-level. Such symmetries have been proposed in the past as a way to maintain a hierarchically small Higgs mass and also play a role in schemes of cosmological relaxation. It is argued that on general grounds one expects shift-symmetries to be restored in the limit of certain asymmetric compactifications, to all orders in perturbation theory. This behaviour is verified by explicit computation of the K\"ahler potential to one-loop order. }
\end{abstract}
\vspace{2in}

\vfill
\smallskip}
\end{titlepage}

\onehalfspacing
\section{Introduction}

An interesting property of the effective field theories that emerge from string theory is that they often possess non-compact shift-symmetries. These are symmetries under which two fields, $B$ and $C$ say, transform as $B\rightarrow B+c$, $C\rightarrow C-\bar{c}$. The K\"ahler potential of a theory with such a symmetry, written as a power series expansion in the matter fields, has to take the form 
\begin{equation}\label{SSKP}
K=G+\abs{B+\bar{C}}^2 f +\ldots \, ,
\end{equation} 
where the coefficients $G$ and $f$ will generally have some dependence on the K\"{a}hler and complex structure moduli of the compactification.
Consequently the orthogonal combination $B-\bar{C}$ remains massless. An observation made by \cite{Cardoso1,EffectiveMu,earlyshift,earlyshift2} and discussed further in \cite{SSHeb}, is that these seemingly ad-hoc continuous symmetries appear naturally at tree-level due to the underlying discrete modular symmetries of the full string theory. They were initially suggested as a way of directly protecting Higgs masses. Furthermore it has been observed that shift symmetries may be linked to the apparent vanishing of the Higgs self-coupling at intermediate scales~\cite{SSHeb}. 

It is an unfortunate fact that the shift-symmetries in question are only accidental and global. One does not expect them to be preserved, even at the string scale, because the full string theory does not respect them. Nevertheless an interesting question is how quickly such symmetries are eroded in perturbation theory, and whether there is a parametric way of controlling them or possibly even restoring them in the string thresholds. Although there has been some work done on one-loop corrections to the effective $\mu$-term for example \cite{EffectiveMu}, this particular issue has not to our knowledge been explored in any detail. 

Although it is a generic expectation that non-compact shift-symmetries afford no more than a loop's worth of protection for any would-be Higgs field, the purpose of this paper is to show that in the limit of certain asymmetric compactifications the symmetries are preserved. Indeed they can be made parametrically good at the string scale. 

There is a simple general argument supporting the restoration of shift-symmetries in asymmetric compactifications which is as follows. Consider the class of heterotic string theories that exhibit $\mathcal{N}=1$ supergravity as their low energy effective field theories, and have a $\mathbb{T}_2/\mathbb{Z}_2$ orbifold subfactor in their compactification (although almost certainly the heuristic argument we are about to present applies more generally). The K\"ahler and complex structure moduli of the $\mathbb{T}_2/\mathbb{Z}_2$ are denoted $T,U$.
We will consider our theory in the presence of two continuous Wilson lines associated with each of the two compact dimensions of the $T^2$, 
a linear combination of which corresponds to the matter fields $B$ and $C$.
For the untwisted components we are then interested in whether the coefficients $H_{\scriptscriptstyle{BC}}(T,U)$, $Z_{\scriptscriptstyle{B\bar{B}}}(T,U)$ and $Z_{\scriptscriptstyle{C\bar{C}}}(T,U)$ in 
\begin{equation}\label{KPExp}
K=G+Z_{\scriptscriptstyle{C\bar{C}}}C{\bar{C}}+Z_{\scriptscriptstyle{B\bar{B}}}B {\bar{B}}+(H_{\scriptscriptstyle{BC}}C B +\text{c.c.})+\ldots,
\end{equation}
  exhibit the correct relation at one-loop order, so that it can be cast in the form \eqref{SSKP}.

At tree-level, the K\"{a}hler potential is well known for such models, and is given by \cite{Cardoso1,EffectiveMu},
\begin{equation}
K=-\log\left[-(T-\bar{T})(U-\bar{U})-(B+\bar{C})(\bar{B}+C)\right],
\end{equation}
clearly exhibiting the shift-symmetry in question. To see why we expect the shift-symmetry to be preserved at higher order in certain limits, 
we recall the particular linear combination 
of complex Wilson lines $A^1$ and $A^2$ (where upstairs indices label two different Cartan subalgebra $U(1)$'s) giving rise to
 $B$ and $C$:
\begin{equation}
\label{bdef}
B=-\frac{1}{\sqrt{2}}(iA^1+A^2),
\quad
C=-\frac{1}{\sqrt{2}}(iA^1-A^2).
\end{equation}
These are each further related to two real Wilson lines as $A^a=U\mathcal{A}_1^a-\mathcal{A}_2^a$, where the lower indices label the two $\mathbb{T}_2$ cycles). The real Wilson lines represent shifts in the internal momentum/charge lattice (a.k.a. Narain lattice) of the compactification, so they can be thought of as directly corresponding to the original stringy degrees of freedom. The crucial point is that in the highly asymmetric ($U_2\gg 1$) limit, $A^a$ is dominated by the term $iU_2\mathcal{A}_1^a$, where in our convention $U=U_1 + i U_2$. Comparing the expressions for $\bar{B}$ and $C$ in this limit, we see that they are both given by,
\begin{equation}
\bar{B},C = \frac{U_2}{2}(\mathcal{A}_1^1+i\mathcal{A}_1^2) + {\cal O}(1).
\end{equation}
Not surprisingly at large $U_2$ the two Wilson lines are both dominated by one of the cycles and they become degenerate. 
The general expectation therefore is that all radiative corrections to the K\"ahler potential exhibit degeneracy for $B$ and $C$ in the limit 
of large $U_2$. In particular one would naturally expect the coefficients of $B\bar{B}$ and $BC$ to become degenerate {\it to all orders}. 

We would like to test this heuristic expectation, and in order to do so we will compute the relevant corrections to the K\"{a}hler potential at one-loop, allowing us to determine and study the coefficients  $H_{\scriptscriptstyle{BC}}(T,U)$, $Z_{\scriptscriptstyle{B\bar{B}}}(T,U)$ and $Z_{\scriptscriptstyle{C\bar{C}}}(T,U)$ appearing in \eqref{KPExp}.
It will be sufficient to find the one-loop corrections to the K\"{a}hler potential up to quadratic order in the untwisted matter fields. Therefore we will proceed by computing the CP even part of one-loop two-point functions involving the moduli $T$ and $U$ as the external states but with the continuous Wilson line moduli in place. We can then focus on the $\mathcal{O}(k^2)$ piece of the amplitude, and compare it with the corresponding kinetic terms in the effective supergravity Lagrangian. Those terms are of the form,
\begin{equation}
K_{i\bar{j}}\partial\phi^i\partial\phi^{\bar{j}},
\end{equation}
so essentially it is the K\"{a}hler metric $K_{i\bar{j}}$ that we compute, from which one could then hope to determine the K\"{a}hler potential. This method was utilised in \cite{Berg1} to calculate one-loop corrections to the K\"{a}hler potential for type-II strings compactified on orientifolds, and a similar procedure was also performed for heterotic strings in \cite{FS1}. Furthermore, loop corrections to low-energy effective theories of heterotic strings have also been investigated in \cite{HEFT}.

The bulk of the computation is carried out in the next section: we first introduce the notation for the moduli and partition function in the presence of Wilson lines, and then consider the two-point amplitude between moduli $T$ and $\bar{T}$, evaluating the relevant correlation functions. Then we compute the integrals over $\tau$ by the unfolding method. In section 3 we use the results to write a consistent expression for the one-loop corrections to the K\"{a}hler potential up to quadratic order in the Wilson lines, and confirm the general picture outlined above. 
Indeed in theories of this kind we find that $\varepsilon=1/(T_2+U_2)$ is a small parameter governing shift-symmetry violation in the limit 
that $U_2\gg 1$, while conversely when $U_2 \sim 1$ there is no shift-symmetry at all in the effective theory at the string scale\footnote{Note that there is {\it no-scale} symmetry which sets all the relevant scalar masses zero at tree-level, but shift-symmetry itself is absent.}.

\section{The calculation} 
\subsection{Moduli definitions, vertex operators and partition function}
\label{sec:Notation}
Let us begin by gathering some necessary ingredients. As per the introduction, we will focus on models where the compactification includes an orbifolded two-torus, and focus on the contributions that arise due to the presence of the two real non-zero Wilson lines $\mathcal{A}_1^a$ and $\mathcal{A}_2^a$. These are mixed with the K\"ahler and complex structure moduli in their relation to the metric and antisymmetric tensor; the required relation is \cite{Cardoso1,KiritsisObers}
\begin{equation}
T=i\sqrt{G}+B_{12}+\frac{1}{2}\sum_aA^a\frac{A^a-\bar{A}^a}{U-\bar{U}},
\end{equation}
where, as above, the complex Wilson lines are defined as $A^a=U\mathcal{A}_1^a-\mathcal{A}_2^a$. 
The $U$ modulus is unchanged by the presence of Wilson lines and so it can simply be defined in the usual way as,
\begin{equation}
U=\frac{1}{G_{11}}\left(i\sqrt{G}+G_{12}\right).
\end{equation}
From the above, we can then write the metric $G_{IJ}$ and antisymmetric tensor $B_{IJ}$ for the torus as follows,
\begin{equation}
G_{IJ}=\left(\frac{T-\bar{T}}{U-\bar{U}}-\frac{(A^a-\bar{A}^a)^2}{2(U-\bar{U})^2}\right)
\begin{pmatrix}
1 && U_1
\\
U_1 && \abs{U}^2
\end{pmatrix},
\end{equation}
\begin{equation}
B_{IJ}=\left(\frac{T+\bar{T}}{2}-\frac{(A^a-\bar{A}^a)(A^a+\bar{A}^a)}{4(U-\bar{U})}\right)
\begin{pmatrix}
0 && 1
\\
-1 && 0
\end{pmatrix}.
\end{equation}

The specific calculation we will perform is the two-point function between the moduli $T$ and $\bar{T}$,  so next we need the corresponding vertex operators. In terms of real coordinates, the vertex operators for the moduli in the zero picture are given by \cite{KiritsisBook,AGN},
\begin{equation}
V_{T^i}=v_{IJ}^{(T^i)}:(\partial X^I+ik\cdot\psi\psi^I)\bar{\partial}X^Je^{ik\cdot X}:,
\end{equation}
where $T^i$ denotes both the moduli $T$ and $U$, and,
\begin{equation}
v_{IJ}^{(T^i)}=\frac{\partial}{\partial T^i}(G_{IJ}+B_{IJ}).
\end{equation}
We find it more convenient to use a similar notation to \cite{Berg1}, and to write the vertex operators in terms of the complex coordinates $Z$ and $\Psi$ defined as,
\begin{equation}
\begin{alignedat}{4}
Z&=\sqrt{\frac{T_2+\frac{(A-\bar{A})^2}{8U_2}}{2U_2}}(X^5+\bar{U}X^6), \qquad & \bar{Z}&=\sqrt{\frac{T_2+\frac{(A-\bar{A})^2}{8U_2}}{2U_2}}(X^5+UX^6),
\\
\Psi&=\sqrt{\frac{T_2+\frac{(A-\bar{A})^2}{8U_2}}{2U_2}}(\psi^5+\bar{U}\psi^6), \qquad  &\bar{\Psi}&=\sqrt{\frac{T_2+\frac{(A-\bar{A})^2}{8U_2}}{2U_2}}(\psi^5+U\psi^6).
\end{alignedat}
\end{equation}
The vertex operator for the $T$ modulus can then be written in the zero picture as,
\begin{equation}
V_T=-\frac{i}{T_2+\frac{(A-\bar{A})^2}{8U_2}}(\partial Z-ik\cdot\psi\Psi)\bar{\partial}\bar{Z}e^{ik\cdot X},
\end{equation}
while for the $U$ modulus we have,
\begin{equation}
V_U=-\frac{i(A-\bar{A})^2}{8U_2^2\left(T_2+\frac{(A-\bar{A})^2}{8U_2}\right)}(\partial Z-ik\cdot\psi\Psi)\bar{\partial}\bar{Z}e^{ik\cdot X}+\frac{i}{U_2}(\partial Z-ik\cdot\psi\Psi)\bar{\partial}Ze^{ik\cdot X}.
\end{equation}

We shall also need the internal partition function associated with the torus. With the inclusion of the Wilson lines, the relevant contribution can be written as \cite{KiritsisObers},
\begin{equation}
{\cal Z}_{\vec{m},\vec{n}}(T,U,\vec{\mathcal{A}}^a)=\frac{T_2+\frac{(A-\bar{A})^2}{8U_2}}{\tau_2}\sum_{\vec{m},\vec{n}\in\mathbb{Z}}e^{-S(\vec{m},\vec{n})}\sum_{Q^a}q^{(Q^a+\vec{\mathcal{A}}^a\cdot\vec{n})^2/2}e^{-2\pi i\vec{\mathcal{A}}^a\cdot \vec{m}(Q^a+\vec{\mathcal{A}}^a\cdot \vec{n}/2)},
\end{equation}
where,
\begin{equation}
S(\vec{m},\vec{n})=\frac{\pi}{\tau_2}(G_{IJ}+B_{IJ})(m_I+n_I\tau)(m_J+n_J\bar{\tau})\, ,
\end{equation}
and $Q^a$ are the elements of the charge/momentum lattice on the gauge side that are shifted by the Wilson lines. Hence only $q$ appears here: the full partition function includes an additional factor we shall refer to as ${\cal Z}_{rest}(q,\bar{q}) $  that is unshifted by the Wilson lines, which incorporates the remaining degrees of freedom (for example those coming from the remaining $K_3$ factor in the compactification). 

\subsection{Two-point Amplitudes}
As previously mentioned, we will obtain the one-loop corrections to the K\"{a}hler potential by computing one-loop amplitudes between the various modulus and anti-modulus pairs,  specifically those corresponding to corrections to $K_{T_i\bar{T}_j}$. This will then allow us to determine the form of the K\"{a}hler potential itself. The amplitudes we need are therefore of the form,
\begin{equation}
\int_{\mathcal{F}}\frac{d^2\tau}{\tau_2^2}\int d^2z\langle V_{T_i}(k,z)V_{\bar{T}_j}(-k,0)\rangle {\cal Z}_{\vec{m},\vec{n}}{\cal Z}_{\text{rest}}.
\end{equation}
The correlation function between the  vertex operators is
\begin{equation}
\langle V_TV_{\bar{T}}\rangle=-\frac{1}{\left(T_2+\frac{(A-\bar{A})^2}{8U_2}\right)^2}\langle(\partial Z-ik\cdot\psi\Psi)\bar{\partial}\bar{Z}e^{ik\cdot X}(\partial \bar{Z}+ik\cdot\psi\bar{\Psi})\bar{\partial}Ze^{-ik\cdot X}\rangle.
\end{equation}
In a supersymmetric theory, the only non-zero contribution to the amplitude arises when all four of the fermionic coordinates are contracted, because the remaining pieces are spin independent and will therefore vanish by the non-renormalisation theorem (i.e. they get multiplied by the partition function which is zero). 
 Even in non-supersymmetric theories, as in \cite{NonSUSY}, the remaining pieces would 
 be proportional to the cosmological constant and hence suppressed if the latter is suppressed. Of course the vanishing of the cosmological constant beyond one-loop in such theories is very much still under investigation and so the stability of such models can not be guaranteed. Nevertheless, for the models under consideration we need only consider the spin dependent term,
	\begin{equation}
	-\frac{1}{4\left(T_2+\frac{(A-\bar{A})^2}{8U_2}\right)^2}k^2\langle\psi\cdot\psi\rangle\langle\Psi\bar{\Psi}\rangle\langle\bar{\partial}\bar{Z}\bar{\partial}Z\rangle.
	\end{equation}
For the bosonic correlation function we will only need to consider the contributions arising from the zero-modes, for which we have,
\begin{equation}
\langle\bar{\partial}Z(z)\bar{\partial}\bar{Z}(0)\rangle=\sum_{\vec{m},\vec{n}}\frac{\pi^2\left(T_2+\frac{(A-\bar{A})^2}{8U_2}\right)}{\tau_2^2U_2}[m_1+n_1\bar{\tau}+U(m_2+n_2\bar{\tau})][m_1+n_1\bar{\tau}+\bar{U}(m_2+n_2\bar{\tau})].
\end{equation}
Given the lack of $z$-dependence in the above, in order to compute the integral over $z$ we need only take into account the contributions from the fermionic correlation functions. The integral is calculated as in \cite{NonSUSY}:
\begin{align}
\begin{split}
I &= \int d^2z\langle\psi^\rho\psi^\sigma\rangle \langle\Psi\bar{\Psi}\rangle \label{zintegral}
\\
&= \int d^2z\left(\wp+4\pi i\partial_{\tau}\log\sqrt{\vartheta_{ab}(0)\vartheta_{cd}(0)}/\eta(\tau)\right)
\\
&= \int d^2z\left(-\partial_{z}^2\log\vartheta_1(z)+4\pi i\partial_{\tau}\log\sqrt{\vartheta_{ab}(0)\vartheta_{cd}(0)}\right)
\\
&=\pi+4\pi i\tau_2 \partial_{\tau}\log\sqrt{\vartheta_{ab}(0)\vartheta_{cd}(0)}\, ,
\end{split}
\end{align}
where $a,b$ and $c,d$ refer to the spin structures of $\psi$ and $\Psi$ respectively, which is being summed over. Note that, analogously to the usual beta function calculation, the second term can also be written as $2\pi i \partial_\tau(Z_\psi Z_{\Psi})$. 
Here, we can now take note of the fact that our amplitude includes a sum over all of the spin structures. The spin independent contribution therefore vanishes after the sum is taken, and so we are left only with the term proportional to $\tau_2$. 

What remains is to calculate is the following integral,
\begin{equation}
\frac{-\pi^2k^2}{4\left(T_2+\frac{(A-\bar{A})^2}{8U_2}\right)U_2}\int_{\mathcal{F}}\frac{d^2\tau}{\tau_2^3}\sum_{\vec{m},\vec{n}}[m_1+n_1\bar{\tau}+U(m_2+n_2\bar{\tau})][m_1+n_1\bar{\tau}+\bar{U}(m_2+n_2\bar{\tau})]{\cal Z}_{\vec{m},\vec{n}}\tilde{{\cal Z}}_{\text{rest}}.
\end{equation}
where now $\tilde{{\cal Z}}_{\text{rest}}$ is given by ${\cal Z}_{\text{rest}}$ with the inclusion of the extra spin dependent piece from the fermion correlators as given by  \eqref{zintegral}. Note that the factor of $\tau_2$ has already been extracted from this additional piece, and $\tilde{{\cal Z}}_{\text{rest}}$ also contains the sum over spin structures.
We now proceed to expand this expression in terms of the Wilson lines. We can then focus on the quadratic terms, and subsequently evaluate the corresponding integrals.

\subsection{Modular Integrals}
In order to compute the modular integrals arising from the two-point functions, we can use the unfolding technique of \cite{DKL1} (also utilised in \cite{FS1,NonW,HetTypeI}), in which
the integral is split into representative orbits of $SL(2,\mathbb{Z})$. This decomposes the integral over the fundamental domain into simpler integration regions, depending on the type of orbit. There are three types of orbits, the zero orbit, degenerate orbits and non-degenerate orbits. We begin by writing the partition function in terms of complex Wilson lines in the form \cite{HarMoore,KiritsisObers}
\begin{equation}
{\cal Z}_{\vec{m},\vec{n}}(T,U,\vec{\mathcal{A}}^a)=\frac{T_2+\frac{(A-\bar{A})^2}{8U_2}}{\tau_2}\sum_{Q^a}q^{Q\cdot Q/2}e^{\mathcal{G}(M,\tau)},
\end{equation}
where
\begin{align}
\begin{split}
\mathcal{G}(M,\tau)=&\frac{-\pi\left(T_2+\frac{(A-\bar{A})^2}{8U_2}\right)}{\tau_2U_2}\abs{\mathcal{M}}^2-2\pi iT\det M+\frac{\pi}{U_2}\left(Q\cdot A\,\tilde{\mathcal{M}}-Q\cdot\bar{A}\,\mathcal{M}\right)
\\
&-\frac{\pi n_2}{2U_2}\left(A\cdot A\,\tilde{\mathcal{M}}-\bar{A}\cdot\bar{A}\,\mathcal{M}\right)-\frac{i\pi(A-\bar{A})^2}{4U_2^2}(n_1+n_2\bar{U})\mathcal{M},
\end{split}
\end{align}
and
\begin{equation}
M=
\begin{pmatrix}
n_1 & m_1
\\
n_2 & m_2
\end{pmatrix},\quad
\mathcal{M}=
\begin{pmatrix}
1 & U
\end{pmatrix}
M
\begin{pmatrix}
\tau
\\
1
\end{pmatrix},\quad
\tilde{\mathcal{M}}=
\begin{pmatrix}
1 & \bar{U}
\end{pmatrix}
M
\begin{pmatrix}
\tau
\\
1
\end{pmatrix}.
\end{equation}
The orbits of $SL(2,\mathbb{Z})$ are then defined in terms of the matrix $M$.
\subsubsection*{Zero Orbit}
This orbit consists only of the matrix $M=0$, with the integration being performed over the fundamental domain. However its contribution trivially vanishes due to the presence of the overall factor from the bosonic zero modes.

\subsubsection*{Degenerate Orbits}
These consist of matrices of the form,
\[
M=
\begin{pmatrix}
	0 && j
	\\
	0 && p
\end{pmatrix},
\]
where the sum is over all integer values $(j,p)\neq(0,0)$ and the integration is extended from the fundamental domain to the half-strip, $E=\{-\frac{1}{2}<\tau_1<\frac{1}{2},\ \tau_2>0\}$. The integral we need to evaluate is of the form,
\begin{equation}\label{DegInt}
\mathcal{I}_1=\frac{-\pi^2}{4\left(T_2+\frac{(A-\bar{A})^2}{8U_2}\right)U_2}\int_E\frac{d^2\tau}{\tau_2^3}\sum_{(j,p)\neq(0,0)}\abs{j+pU}^2{\cal Z}_{(j,p),(0,0)}\tilde{{\cal Z}}_{\text{rest}},
\end{equation}
where the partition function becomes,
\begin{align}
\begin{split}
{\cal Z}_{(j,p),(0,0)}=&\frac{T_2+\frac{(A-\bar{A})^2}{8U_2}}{\tau_2}\exp\left[-\frac{\pi}{\tau_2U_2}\left(T_2+\frac{(A-\bar{A})^2}{8U_2}\right)\abs{j+pU}^2\right]
\\
&\times\sum_{Q^a}q^{Q\cdot Q/2}\exp\left[\frac{\pi}{U_2}\left[Q\cdot A(j+p\bar{U})-Q\cdot\bar{A}(j+pU)\right]\right].
\end{split}
\end{align}
As mentioned, we are primarily interested in calculating the K\"{a}hler potential only up to quadratic order in the Wilson lines. Therefore, we can write the above as an expansion in  $A^a$ and $\bar{A}^a$, and focus only on the relevant terms. 

To begin, we can evaluate the Wilson line independent part of \eqref{DegInt}:
\begin{align}
\begin{split}
\frac{-\pi^2}{4U_2}\int_E\frac{d^2\tau}{\tau_2^4}\sum_{\substack{(j,p)\neq(0,0)\\Q^a}}\abs{j+pU}^2e^{-\frac{\pi T_2}{\tau_2U_2}\abs{j+pU}^2}q^{Q\cdot Q/2}\tilde{{\cal Z}}_{\text{rest}}
&= c_1\frac{4i}{\pi(T-\bar{T})^3}E(U,2)+\ldots\ ,
\end{split}
\end{align}
where we have written only the most dominant contribution, and $c_1$ is some constant of order one that we do not calculate. It is dependent on the coefficients of the power series in $q$ and $\bar{q}$ in $q^{Q\cdot Q/2}\tilde{{\cal Z}}_{\text{rest}}$, the sum over spin structures, and also on a restricted sum over the lattice vectors $Q^a$.
In the above, the real analytic Eisenstein series are defined as,
\begin{equation}
E(U,s)=\sideset{}{'}\sum_{l,m}\frac{U_2^s}{|l+mU|^{2s}},
\end{equation}
where the prime means we do not include the case when $l_1=l_2=0$ in the sum. 

We now extract the terms proportional to $A^a\bar{A}^a$ and $A^aA^a$. The former term is given by,
\begin{equation}\label{IAA}
\mathcal{I}_1^{A,\bar{A}}=\frac{-\pi^3}{4U_2^3}\int_E\frac{d^2\tau}{\tau_2^4}\sum_{\substack{(j,p)\neq(0,0)\\Q^a}}F(A,\bar{A})\abs{j+pU}^2e^{-\frac{\pi T_2}{\tau_2U_2}\abs{j+pU}^2}q^{Q\cdot Q/2}\tilde{{\cal Z}}_{\text{rest}},
\end{equation}
where
\begin{equation}
F(A,\bar{A})=\left(\frac{1}{4\tau_2}A^a\bar{A}^a-\pi (Q\cdot A)(Q\cdot\bar{A})\right)\abs{j+pU}^2.
\end{equation}
The integral over $\tau$ can be performed with the result
\begin{equation}\label{DegBC}
\tilde{\mathcal{I}}_1^{A,\bar{A}}=\frac{-12ic_1 E(U,2)}{\pi(T-\bar{T})^4(U-\bar{U})}+\frac{4\pi^2c_2}{(T-\bar{T})^3(U-\bar{U})}\left[3-2\log(-e^{-2\gamma}\pi(T-\bar{T})(U-\bar{U})\abs{\eta(U)}^4)\right],
\end{equation}
where $\gamma$ is the Euler-Mascheroni constant and $\mathcal{I}_1^{A,\bar{A}}=\tilde{\mathcal{I}}_1^{A,\bar{A}}A\bar{A}$. Note that in order to arrive at the above result it is necessary to regulate the divergent parts of the integral (proportional to $\tau_2^{-4}$ in the integrand) that have arisen because we have exchanged the order of summation and integration. These can be dealt with by including an additional factor of $\tau_2^{-\epsilon}$, performing the integration, evaluating the sum and extracting the $\epsilon$ independent piece as described in \cite{FS2,RegV}. Alternatively, one finds the same result using the regularisation procedure of \cite{DKL1}. As before, the constants $c_1$ and $c_2$ come from the coefficients of the power series in $q$ and $\bar{q}$ in ${\cal Z}_{\text{rest}}$, the sum over spin structures, and from the sum over lattice vectors $Q^a$; they are completely independent of moduli.

Similarly, the expression we need  for the term proportional to $A^aA^a$ is,
\begin{equation}
\mathcal{I}_1^{A,A}=\frac{-\pi^3}{4U_2^3}\int_E\frac{d^2\tau}{\tau_2^4}\sum_{\substack{(j,p)\neq(0,0)\\Q^a}}F(A,A)\abs{j+pU}^2e^{-\frac{\pi T_2}{\tau_2U_2}\abs{j+pU}^2}q^{Q\cdot Q/2}\tilde{{\cal Z}}_{\text{rest}},
\end{equation}
\begin{equation}
F(A,A)=\left(-\frac{1}{8\tau_2}\abs{j+pU}^2A^aA^a+\frac{\pi}{2}(j+p\bar{U})^2(Q\cdot A)^2\right)\, ,
\end{equation}
where again the integral over $\tau$ can be performed with suitable regularisation and we obtain the result,
\begin{equation}
\tilde{\mathcal{I}}_1^{A,A}=\frac{6ic_1 E(U,2)}{\pi(T-\bar{T})^4(U-\bar{U})}+\frac{4\pi^2c_2}{(T-\bar{T})^3}\left[2\partial_U\log\eta(U)+\frac{1}{(U-\bar{U})}\right].
\end{equation}
Finally, the result for the term proportional to $\bar{A}^a\bar{A}^a$ is just given by the complex conjugate of $\tilde{\mathcal{I}}_1^{A,A}$.

\subsubsection*{Non-degenerate Orbits}
These consist of matrices of the form,
\[
M=\pm
\begin{pmatrix}
k && j
\\
0 && p
\end{pmatrix},
\]
where the sum is over $0\leq j<k,\ p\neq0$ and the integration is over the upper half plane $\mathbb{H}$. The expression to evaluate is of the form,
\begin{equation}\label{NonDegInt}
\mathcal{I}_2=\frac{-\pi^2}{4\left(T_2+\frac{(A-\bar{A})^2}{8U_2}\right)U_2}\int_\mathbb{H}\frac{d^2\tau}{\tau_2^4}\sum_{\substack{0\leq j<k\\p\neq 0}}\tilde{Q}_U\tilde{Q}_{\bar{U}}{\cal Z}_{(j,p),(k,0)}\tilde{{\cal Z}}_{\text{rest}},
\end{equation}
where the torus partition function is,
\begin{align}
\begin{split}
{\cal Z}_{(j,p),(k,0)}=&\frac{T_2+\frac{(A-\bar{A})^2}{8U_2}}{\tau_2}\exp\left[-\frac{\pi T_2}{U_2\tau_2}\abs{Q_U}^2-2\pi iTkp-\frac{\pi(A-\bar{A})^2}{8U_2^2\tau_2}\abs{Q_U}^2-\frac{\pi i(A-\bar{A})^2}{4U_2^2}kQ_U\right]
\\
&\times\sum_{Q^a}q^{Q\cdot Q/2}\exp\left[\frac{\pi}{U_2}\left(Q\cdot A\,Q_{\bar{U}}-Q\cdot\bar{A}\,Q_U\right)\right]
\end{split}
\end{align}
and where,
\begin{align}
\begin{split}
Q_U&=(j+k\tau+pU),
\\
Q_{\bar{U}}&=(j+k\tau+p\bar{U}),
\\
\tilde{Q}_U&=(j+k\bar{\tau}+pU),
\\
\tilde{Q}_{\bar{U}}&=(j+k\bar{\tau}+p\bar{U}).
\end{split}
\end{align}

As for the degenerate orbits, we will evaluate the first few terms in a series expansion of \eqref{NonDegInt} in the Wilson lines. The result for the Wilson line independent part (after summing over $j$ and $p$) is,
\begin{align}\label{NonDWLIndep}
\begin{split}
&\frac{-\pi^2}{4U_2}\int_\mathbb{H}\frac{d^2\tau}{\tau_2^4}\sum_{\substack{0\leq j<k\\p\neq 0, \, Q_a}}\tilde{Q}_U\tilde{Q}_{\bar{U}}e^{-2\pi iTkp}e^{-\frac{\pi T_2}{\tau_2U_2}\abs{j+k\tau+pU}^2}q^{Q\cdot Q/2}\tilde{{\cal Z}}_{\text{rest}}
\\
=&\frac{-4c_1}{(T-\bar{T})^3(U-\bar{U})}\sum_{k>0}\left\{2k\pi T_2\left[\Li_2\left(q_T^k\right)+\Li_2\left(\bar{q}_T^k\right)\right]+\left[\Li_3\left(q_T^k\right)+\Li_3\left(\bar{q}_T^k\right)\right]\right\}+\ldots\ ,
\end{split}
\end{align}
where $q_T\equiv\exp(2\pi i T)$ and the polylogarithms $\Li_n(z)$ are defined as,
\begin{equation}
\Li_n(z)=\sum_{k>0}\frac{z^k}{k^n}.
\end{equation} 
In the above we are again only writing the dominant contributions. A more complete expression could be obtained along the lines of \cite{FS1}, but taking only these terms is sufficient for the comparison between the terms $Z$ and $H$ in the K\"{a}hler potential.

Now, as in the case for the degenerate orbits, we can look at the terms proportional to $A^a\bar{A}^a$. These are given by,
\begin{equation}
\mathcal{I}_2^{A,\bar{A}}=\frac{-\pi^3}{8U_2^3}\int_\mathbb{H}\frac{d^2\tau}{\tau_2^4}\sum_{\substack{0\leq j<k\\p\neq 0,\, Q_a}}F(A,\bar{A})
\tilde{Q}_U\tilde{Q}_{\bar{U}}e^{-2\pi iTkp}e^{-\frac{\pi T_2}{\tau_2U_2}\abs{j+k\tau+pU}^2}q^{Q\cdot Q/2}\tilde{{\cal Z}}_{\text{rest}},
\end{equation}
where,
\begin{equation}
F(A,\bar{A})=\left[-2\pi Q_UQ_{\bar{U}}(Q\cdot A)(Q\cdot\bar{A})+\left(ikQ_U+\frac{1}{2\tau_2}\abs{Q_U}^2\right)A^a\bar{A}^a\right].
\end{equation}
Performing the integration over $\tau$ and summing over $j$ and $p$ we obtain the result,
\begin{align}
\begin{split}
\tilde{\mathcal{I}}_2^{A,\bar{A}}=&
\frac{4}{(T-\bar{T})^4(U-\bar{U})^2}\bigg\{c_1\sum_{k>0}\bigg[\pi^2(T-\bar{T})^2k^2\left[\log\left(1-q_T^k\right)+\log\left(1-\bar{q}_T^k\right)\right]
\\
&-3\pi ik(T-\bar{T})\left[\Li_2\left(q_T^k\right)+\Li_2\left(\bar{q}_T^k\right)\right]+3\left[\Li_3\left(q_T^k\right)+\Li_3\left(\bar{q}_T^k\right)\right]\bigg]
\\
&.+\pi^2ic_2(T-\bar{T})^2(U-\bar{U})\left[\partial_T\log\eta(T)-\partial_{\bar{T}}\log\eta(\bar{T})\right]
\\
&-\pi^2c_2(T-\bar{T})(U-\bar{U})\log\abs{\eta(T)}^4\bigg\}.
\end{split}
\end{align}

Moving on to the terms proportional to $A^aA^a$, we wish to calculate,
\begin{equation}
\mathcal{I}_2^{A,A}=\frac{-\pi^3}{16U_2^2}\int_\mathbb{H}\frac{d^2\tau}{\tau_2^4}\sum_{\substack{0\leq j<k\\p\neq 0\, , Q_a}}F(A,A)\tilde{Q}_U\tilde{Q}_{\bar{U}}e^{-2\pi iTkp}e^{-\frac{\pi T_2}{\tau_2U_2}\abs{j+k\tau+pU}^2}q^{Q\cdot Q/2}\tilde{Z}_{\text{rest}},
\end{equation}
where,
\begin{equation}
F(A,A)=\left[2\pi Q_{\bar{U}}^2(Q\cdot A)(Q\cdot A)-\left(ikQ_U+\frac{1}{2\tau_2}\abs{Q_U}^2\right)A^aA^a\right].
\end{equation}
Again, computing the integration over $\tau$ and summing over $j$ and $p$, we have the result,
\begin{align}
\begin{split}
\tilde{\mathcal{I}}_2^{A,A}=&
\frac{-2c_1}{(T-\bar{T})^4(U-\bar{U})^2}\sum_{k>0}\left\{\pi^2(T-\bar{T})^2k^2\left[\log\left(1-q_T^k\right)+\log\left(1-\bar{q}_T^k\right)\right]\right.
\\
&\left.-3\pi ik(T-\bar{T})\left[\Li_2\left(q_T^k\right)+\Li_2\left(\bar{q}_T^k\right)\right]+3\left[\Li_3\left(q_T^k\right)+\Li_3\left(\bar{q}_T^k\right)\right]\right\}.
\end{split}
\end{align}

\section{One-loop K\"{a}hler potential}
From the results of the previous section it is possible to establish the form of the one-loop corrections to the K\"{a}hler potential. In order to compare them to the corresponding kinetic terms in the supergravity Lagrangian, we Weyl rescale to the Einstein frame giving an additional factor\,
\begin{equation}
e^{2\Phi}=\frac{2i}{S-\bar{S}}.
\end{equation}
We wish to express the K\"{a}hler potential in the form in \eqref{KPExp},with the Wilson lines and their complex conjugates defined as in \eqref{bdef}.
Taking the sum over the index $a$ we find, $\sum_aA^a\bar{A}^a=B\bar{B}+C\bar{C}$, and the one-loop corrections to the coefficients $Z_{B\bar{B}}$ and $Z_{C\bar{C}}$ both then satisfy,
\begin{equation}\label{Z1Diff}
\partial_T\partial_{\bar{T}}Z^{(1)}=\frac{2i}{S-\bar{S}}\left(\tilde{\mathcal{I}}_1^{A,\bar{A}}+\tilde{\mathcal{I}}_2^{A,\bar{A}}\right),
\end{equation}
where $\tilde{\mathcal{I}}_1^{A,\bar{A}}$ and $\tilde{\mathcal{I}}_2^{A,\bar{A}}$ are the contributions from the degenerate and non-degenerate orbits respectively, as computed in the previous section.

Similarly, using $\sum_aA^aA^a=-2BC$,  the one-loop correction to the coefficient $H_{BC}$ in \eqref{KPExp} (where again we perform a Weyl rescaling)  satisfies,
\begin{equation}\label{H1Diff}
\partial_T\partial_{\bar{T}}H_{BC}^{(1)}=\frac{-4i}{S-\bar{S}}\left(\tilde{\mathcal{I}}_1^{A,A}+\tilde{\mathcal{I}}_2^{A,A}\right).
\end{equation}

 An additional constraint for the K\"ahler potential that gives the above K\"ahler metric terms is of course that it is required to be invariant under modular transformations of the moduli, up to K\"{a}hler transformations. Taking all of this into account, we find,
	\begin{align}
	\begin{split}
	Z^{(1)}=&\frac{-2c_1 }{\pi(S-\bar{S})(T-\bar{T})(U-\bar{U})}\bigg\{\left(\frac{E(U,2)}{(T-\bar{T})}+\frac{\mathcal{P}(T)}{(U-\bar{U})}\right)\bigg\} \label{KPZ1}
	\\
	&-\frac{4\pi^2c_2 }{(S-\bar{S})(T-\bar{T})(U-\bar{U})}\log\left[-e^{-2\gamma}\pi(T-\bar{T})(U-\bar{U})\abs{\eta(T)\eta(U)}^4\right],
	\end{split}
	\end{align}
	\begin{align}
	\begin{split}
	H_{BC}^{(1)}=&\frac{-2c_1 }{\pi(S-\bar{S})(T-\bar{T})(U-\bar{U})}\bigg\{\left(\frac{E(U,2)}{(T-\bar{T})}+\frac{\mathcal{P}(T)}{(U-\bar{U})}\right)\bigg\}
	\\
	&-\frac{4\pi^2c_2 }{(S-\bar{S})}\bigg\{\frac{\pi^2}{36}+\left[2\partial_U\log\eta(U)+\frac{1}{(U-\bar{U})}\right]\left[2\partial_T\log\eta(T)+\frac{1}{(T-\bar{T})}\right]\bigg\}, \label{KPH1}
	\end{split}
	\end{align}
	where
\begin{equation}
\mathcal{P}(T)=2\pi^2\sum_{m>0}m\left[\Li_2(q_T^m)+\Li_2(\bar{q}_T^m)\right]+\frac{\pi}{T_2}\sum_{m>0}\left[\Li_3(q_T^m)+\Li_3(\bar{q}_T^m)\right]\, .
\end{equation}
The above expressions for $Z^{(1)}$ and $H_{BC}^{(1)}$ can also be shown to be consistent with the other two point amplitudes involving $U$ and $\bar{U}$ or $T$ and $\bar{U}$.

\section{Restoration of shift-symmetry}
Let us now return to our goal, which is to compare the coefficients $Z^{(1)}$ and $H_{BC}^{(1)}$ 
in order to determine whether the shift-symmetry holds at one loop. Were this symmetry to be exact at this order, one would find equal $Z^{(1)}$ and $H_{BC}^{(1)}$. However, only the first lines of  \eqref{KPZ1} and \eqref{KPH1} are explicitly equal. Note also that at 
large $T_2$ these terms are actually sub-leading. Therefore further examination of the remaining terms is required to determine the extent of the breaking of shift-symmetry. These terms can be expressed respectively as,
\begin{align}
\begin{split}
\tilde{Z}=&\frac{-4\pi^2c_2}{(S-\bar{S})(T-\bar{T})(U-\bar{U})}\bigg\{\log[-e^{-2\gamma}\pi(T-\bar{T})(U-\bar{U})]
\\
&+2\sum_{k>0}\left[\log(1-q_U^k)+\log(1-\bar{q}_U^k)\right]+2\sum_{k>0}\left[\log(1-q_T^k)+\log(1-\bar{q}_T^k)\right]\bigg\}
\\
&-\frac{4\pi^2c_2}{(S-\bar{S})}\bigg\{\frac{\pi}{12U_2}+\frac{\pi}{12T_2}\bigg\},
\end{split}
\end{align}
\begin{align}
\begin{split}
\tilde{H}=&\frac{-4\pi^2c_2}{(S-\bar{S})}\bigg\{\frac{2\pi^2}{3}\sum_{k>0}\left[\frac{kq_T^k}{1-q_T^k}+\frac{kq_U^k}{1-q_U^k}\right]-16\pi^2\sum_{k>0}\frac{kq_T^k}{1-q_T^k}\sum_{m>0}\frac{mq_U^m}{1-q_U^m}
\\
&+2\pi\sum_{k>0}\left[\frac{1}{U_2}\frac{kq_T^k}{1-q_T^k}+\frac{1}{T_2}\frac{kq_U^k}{1-q_U^k}\right]-\frac{1}{4T_2U_2}\bigg\}
-\frac{4\pi^2c_2}{(S-\bar{S})}\bigg\{\frac{\pi}{12U_2}+\frac{\pi}{12T_2}\bigg\}.
\end{split}
\end{align}
Aside from the final terms appearing in each of the above expressions, $\tilde{Z}$ and $\tilde{H}$ are not equivalent in general, and so the shift-symmetry will not generically hold. Nevertheless, we are interested in the possibility that in the large $U_2$ limit the shift-symmetry is restored as discussed in the introduction. Any breaking of shift symmetry translates directly into 
shifts in the typical induced soft-terms of the form
\begin{equation}
\frac{\delta m^2}{m^2}=\frac{\Re (\tilde{H}-\tilde{Z})}{Z^{(1)}},
\end{equation}
where $m^2$ is the mass-squared of the heavy Wilson line scalar.
Note that in writing this expression we are using the fact that the {\it tree-level} masses of {\it all} the scalars are zero in these 
theories due to their no-scale structure. Therefore the expression above incorporates the leading one-loop contribution proportional to the gravitino mass $m_{3/2}$. We should also remark that additional contributions to masses come from other one-loop effects such as the Green-Schwarz mechanism, if there is one operating in the theory. Moreover what we are calculating here are stringy thresholds and there will be contributions 
from lighter modes such as stops in a complete model. Of course if one could construct a completely phenomenologically accurate broken MSSM within the string theory one would be able to compute such effects within the string theory as well; so we are focussing on the violations of shift-symmetry that are certain to exist in the string thresholds of any theory of this type.

Let us now test our expectation that this ratio tends to zero in asymmetric compactification; as this implies $T_2\gg 1$, the terms in the K\"{a}hler potential with any dependence on $q_T^k$ are exponentially suppressed, and we can write,
\begin{align}
\begin{split}
\tilde{H}-\tilde{Z}=-\frac{4\pi^2c_2}{(S-\bar{S})}\bigg\{&\frac{2\pi^2}{3}\sum_{k>0}\frac{kq_U^k}{1-q_U^k}+2\pi\sum_{k>0}\frac{1}{T_2}\frac{kq_U^k}{1-q_U^k}+\frac{\log[4\pi e^{-2\gamma}T_2U_2]}{4T_2U_2}
\\
&+\frac{1}{T_2U_2}\sum_{k>0}\left[\log(1-q_U^k)+\log(1-\bar{q}_U^k)\right]-\frac{1}{4T_2U_2}\bigg\},
\end{split}
\end{align}
while for $Z^{(1)}$ we have,
\begin{align}
\begin{split}
Z^{(1)}=&-\frac{ic_1E(U,2) }{4\pi(S-\bar{S})T_2^2U_2}-\frac{4\pi^2c_2}{(S-\bar{S})}\bigg\{\frac{\log[4\pi e^{-2\gamma}T_2U_2]}{4T_2U_2}+\frac{\pi}{12T_2}+\frac{\pi}{12U_2}
\\
&+\frac{1}{T_2U_2}\sum_{k>0}\left[\log(1-q_U^k)+\log(1-\bar{q}_U^k)\right]\bigg\}.
\end{split}
\end{align}
In the limit $U_2\gg 1$, recalling that we also have the condition $T_2>U_2$, we find the dominant contribution to be
\begin{equation}
\frac{\delta m^2}{m^2}\sim\frac{3\log[4\pi e^{-2\gamma}T_2U_2]}{\pi (T_2+U_2)}, \label{dm/m}
\end{equation}
which clearly vanishes in the $T_2 > U_2\rightarrow \infty $ limit as expected, with $1/ (T_2+U_2)$ being the small parameter. 
Conversely, when $T_2\gg1$ but $U_2\ll1$, we find
\begin{equation}
\frac{\delta m^2}{m^2}\sim\frac{4\pi U_2}{3}\sum_{k>0}\frac{kq_U^k}{1-q_U^k},
\end{equation}
which grows as $U_2$ decreases and moreover it is not small.

We should point out that in taking the limits $T_2\rightarrow\infty$ and $U_2\rightarrow\infty$, one needs to be sure that a perturbative computation is still a sensible thing to do. These limits correspond to a large volume theory where the modified loop counting parameter remains small for sufficiently large $S_2=\Im(S)$, in which case a perturbative expansion may still be valid at all energies. 
One-loop threshold corrections imply an upper bound on $T_2$ and $U_2$ \cite{KLGauge}; indeed the loop expansion parameter (essentially the `t~Hooft coupling) is order $T_2/S_2$, implying that large volumes can be achieved with weak string coupling. 

We conclude that ideas such as those presented in ref.\cite{SSHeb} can be extremely effective in highly asymmetric configurations for the general reasons outlined in the Introduction.  Indeed  for the class of compactifications considered here, the heavy Higgs is already one-loop suppressed with respect to the gravitino mass (gaining a mass through RG running as usual in no-scale models), while the light Higgs is further parametrically suppressed by the asymmetry. 
A more model dependent question is of course if and how shift-symmetries are violated by the RG effects of the low energy theory, which may be computed in the effective field-theory as in ref.\cite{SSHeb}. In a complete picture, such violations of shift-symmetry would arise from spontaneous breaking due to for example flavon fields, leading to light pseudo-Nambu-Goldstone modes, which may or may not mix with the Higgs. In principle the techniques presented could be applied to those more complete cases in an entirely stringy setting. 
Here we have seen that even if shift symmetries appear to be a strong feature of the classical field theory, asymmetric compactification is required to protect them in the threshold corrections as well. 

It would of course be useful to consider these questions in more general settings such as constructions involving D-branes in type II, or smooth Calabi-Yaus. Whilst radiative violations of shift-symmetries in the former would almost certainly be calculable (as per \cite{Berg1}) if the backgrounds are sufficiently flat, the latter is notoriously difficult to treat perturbatively. One could hope to develop heuristic arguments along the lines of those in the introduction, and indeed there may be interesting overlaps with shift-symmetry restoration in certain limits of the type II systems in \cite{IIBshift}.
We should remark that shift-symmetries have also come to the fore because of their central role in schemes that try to explain the weak-Planck hierarchy by means of cosmological relaxation \cite{earlyrelaxion}, a subject which has recently received much attention~\cite{relaxion}. Although these often feature axionic (i.e. compact) symmetries, non-compact shift-symmetries may be of more utility given the need for trans-Planckian field excursions. Moreover in supersymmetric theories the two are in any case related by complexification of the Goldstone manifold. Therefore it may be of interest to revisit this question in the present context. 

\subsubsection*{Acknowledgements} We thank Marcus Berg for useful discussions. RJS is funded by an EPSRC studentship.

\end{document}